\documentclass[12pt]{article}
\usepackage{amsmath}
\usepackage{cite}
\usepackage{graphicx}
\usepackage{amsfonts}
\usepackage{amssymb}

\providecommand{\U}[1]{\protect\rule{.1in}{.1in}}

\csname @addtoreset\endcsname{equation}{section}
\newcommand{\eq}{\begin{equation}}
\newcommand{\feq}{\end{equation}}
\newcommand{\eqn}{\begin{eqnarray}}
\newcommand{\feqn}{\end{eqnarray}}
\newcommand{\arr}{\begin{eqnarray*}}
\newcommand{\farr}{\end{eqnarray*}}

\begin{document}

\begin{titlepage}
\begin{center}
\renewcommand{\thefootnote}{\fnsymbol{footnote}}
{\Large{\bf Landau-Lifshitz Solutions of $N=2$, $D=5$ Supergravity}}
\vskip1cm
\vskip 1.3cm
W. A. Sabra
\vskip 0.5cm
{\small{Department of Physics, American University of Beirut, Lebanon }}
\end{center}
\bigskip
\begin{center}
{\bf Abstract}
\end{center}
We study families of solutions depending only on one coordinate for the theories of $N=2$,  five-dimensional
supergravity coupled to vector multiplets. Four families of solutions, valid for all space-time signatures, are obtained. The explicit metrics of the solutions depend on
the Jordan normal forms of constrained matrices. Examples with three charges are given for the so-called STU
model.
\end{titlepage}

\section{INTRODUCTION}

Many years ago, Kasner presented four-dimensional vacuum solutions depending
on one variable \cite{kasner}, which when generalized to arbitrary
space-time signature, take the form \cite{harvey} 
\begin{equation}
ds^{2}=\epsilon _{0}x^{2a_{1}}dx^{2}+\epsilon _{1}x^{2a_{2}}dy^{2}+\epsilon
_{2}x^{2a_{3}}dz^{2}+\epsilon _{3}x^{2a_{4}}dw^{2}  \label{k}
\end{equation}
with the conditions 
\begin{equation}
a_{2}+a_{3}+a_{4}=a_{1}+1,\text{ \ \ \ }a_{2}^{2}+a_{3}^{2}+a_{4}^{2}=\left(
a_{1}+1\right) ^{2}.
\end{equation}
The cosmological synchronous Bianchi I metric \cite{macel} , 
\begin{equation}
ds^{2}=-d\tau ^{2}+\tau ^{2p_{1}}dx^{2}+\tau ^{2p_{2}}dy^{2}+\tau
^{2p_{3}}dz^{2}  \label{g}
\end{equation}
referred to as the Kasner metric, can be obtained from Lorentzian (\ref{k})
through a change of variable or by setting $a_{1}=0$. Here the Kasner
exponents $p_{1}$, $p_{2}$ and $p_{3\text{ }}$satisfy 
\begin{equation}
p_{1}+p_{2}+p_{3}=p_{1}^{2}+p_{2}^{2}+p_{3}^{2}=1\text{ }.
\end{equation}
The metric (\ref{g}) admits generalizations to arbitrary signature $D$
-dimensional gravity, in the form 
\begin{equation}
ds^{2}=\epsilon _{0}d\tau ^{2}+\epsilon _{\mu }\tau ^{2p_{\mu }}\left(
dx^{\mu }\right) ^{2}
\end{equation}
with the exponents $p_{\mu }$ satisfying the conditions 
\begin{equation}
\sum_{\mu =1}^{D-1}p_{\mu }=\sum_{\mu =1}^{D-1}p_{\mu }^{2}=1
\end{equation}
and $\epsilon _{0}$ and $\epsilon _{\mu }$ are constants taking the values $
\pm 1.$

In the analysis of Landau and Lifshitz \cite{landau} of the four-dimensional
gravitational solutions depending on one variable, one starts with the
metric 
\begin{equation}
ds^{2}=\epsilon _{0}d\tau ^{2}+g_{\mu \nu }dx^{\mu }dx^{\nu }
\end{equation}
and obtains, in addition to the Bianchi I metric, non-diagonal vacuum
solutions corresponding to $\omega =\sqrt{\left\vert g\right\vert }=\tau .$ 
\footnote{
The exact solutions of \cite{landau} were also presented in Appendix C of 
\cite{cosmo}} As pointed out in \cite{harvey}, the cases with $a_{1}=-1$
were not considered by Kasner$.$ In the formulation of \cite{landau}, these
missed cases belong to solutions with $\omega =1$.

Starting with Kasner metric and the scalar-Kasner metric \cite{Belinski},
generalized anisotropic Melvin cosmologies and flux tubes were obtained in 
\cite{dowker, kt}. This was achieved using solution generating techniques 
\cite{har, earn}. Generalizations of Kasner metric to theories with form
gauge fields and a coupled dilaton field in all space-time dimensions and
signatures were considered in \cite{s1}. Also generalized Melvin metrics
were found for the theories of $N=2$ supergravity coupled to vector
multiplets in four and five dimensions \cite{s2, s3}. More recently, general
solutions for $D$-dimensional Einstein-Maxwell theory and theories with form
gauge fields and a coupled dilaton, depending on one variable, were obtained
in \cite{new, fs} through an explicit analysis of the equations of motion.

The current work shall deal with the study of general solutions depending on
one variable in $N=2,$ five-dimensional supergravity theories coupled to
vector multiplets. Kasner-like solutions of the five-dimensional
supergravity theory were considered in \cite{s2}. Depending on the choice of
coordinates, the solutions obtained constitute generalizations of
Melvin/Rosen cosmologies, flux-branes and domain walls \cite{melrose}. They
are of importance in addressing questions of gravitational physics and
cosmology \cite{cosmo, book} within supergravity and string theory.

The Lagrangian of our models is given by 
\begin{equation}
\mathcal{L}_{5}=\sqrt{|g|}\left( R-{\frac{\varepsilon }{2}}G_{IJ}\left( \phi
\right) \mathcal{F}_{\mu \nu }^{I}\mathcal{F}^{\mu \nu \,J}-\mathsf{\mathcal{
\ M}}_{ij}\left( \phi \right) \partial _{\mu }\phi ^{i}\partial ^{\mu }\phi
^{j}\right) \text{ }  \label{act}
\end{equation}
where we have ignored a Chern-Simons term which will not be relevant for our
discussions. In these models, we have $n$ scalar fields $\phi ^{i}$ and $n+1$
Abelian gauge fields with field-strengths $\mathcal{F}_{\mu \nu }^{I}$. The $
N=2,$ five-dimensional supergravity theories coupled to vector multiplets,
with Lorentzian signature and $\varepsilon =1$, were first considered by
Gunaydin, Sierra and Townsend in \cite{GST}. These theories can be described
by the so-called very special geometry \cite{superbook}. One defines the
special coordinates $X^{I}=X^{I}(\phi )$ and their duals $X_{I}={\frac{1}{6}}
C_{IJK}X^{J}X^{K},$ which satisfy the conditions 
\begin{equation}
X^{I}X_{I}=1\text{ },\qquad \mathcal{V=}{\frac{1}{6}}
C_{IJK}X^{I}X^{J}X^{K}=1 \text{ }.  \label{cn}
\end{equation}
The constants $C_{IJK}$ are symmetric in all indices. For the
five-dimensional theories obtained via Calabi-Yau compactification \cite{cad}
of eleven-dimensional supergravity \cite{cjs}, $C_{IJK}$ represent the
intersection numbers. The gauge coupling metric $G_{IJ}$ can be expressed in
terms of the special coordinates by 
\begin{equation}
G_{IJ}={\frac{9}{2}}X_{I}X_{J}-{\frac{1}{2}}C_{IJK}X^{K}\text{ }.
\end{equation}
In terms of the scalars $X^{I},$ the $\phi ^{i}$ kinetic term is given by 
\begin{equation}
\mathsf{\mathcal{M}}_{ij}\partial _{\mu }\phi ^{i}\partial ^{\mu }\phi
^{j}=G_{IJ}\partial _{\mu }X^{I}\partial ^{\mu }X^{J}\text{ }.
\end{equation}
We also have the very special geometry relations 
\begin{equation}
G_{IJ}X^{J}={\frac{3}{2}}X_{I}\,,\text{ \ \ }dX_{I}\,=-\frac{2}{3}
G_{IJ}dX^{J},\text{ \ \ }X_{I}dX^{I}\text{\ }=0\text{\ }.  \label{vs}
\end{equation}

Recently, the $N=2,$ five-dimensional supergravity were considered for all
space-time signatures \cite{euclidean, sig, gt}. A class of these
supergravity theories can be obtained via Calabi-Yau 3-folds
compactification of the exotic eleven-dimensional supergravity theories
constructed by Hull \cite{Hull}. Moreover, some of these theories come with
the non-conventional sign of the gauge kinetic terms\ $(\varepsilon=-1)$.

We organize this work as follows. In the next section we analyze the
equations of motion obtained from (\ref{act}). We consider metric solutions
for two different choices for the gauge fields and obtain two sets of
equations characterizing two classes of solutions. In section three, we
present our four different families of solutions. In section four, we
construct explicit examples for the STU model with three independent
charges. We summarize our results in section five.

\section{SOLUTIONS}

The Einstein, Maxwell and scalar field equations of motion derived from (\ref
{act}) take the form 
\begin{align}
R_{\mu \nu }-G_{IJ}\partial _{\mu }X^{I}\partial _{\nu }X^{J}-\varepsilon {G}
_{IJ}\left( \mathcal{F}_{\mu \alpha }^{I}\mathcal{F}_{\nu }^{J}{}^{\alpha }-{
\ \frac{1}{6}}g_{\mu \nu }\mathcal{F}^{2}\right) & =0\text{ },
\label{einstein} \\
\partial _{\mu }\left( \sqrt{|g|}G_{IJ}\mathcal{F}^{I\mu \nu }\right) & =0 
\text{ },  \label{maxwell} \\
\sqrt{|g|}\partial _{i}G_{IJ}\left( \frac{\varepsilon }{2}\mathcal{F}_{\mu
\nu }^{I}\mathcal{F}^{\mu \nu J}+\partial _{\mu }X^{I}\partial ^{\mu
}X^{J}\right) -2\partial _{\mu }\left( \sqrt{|g|}G_{IJ}\partial ^{\mu
}X^{J}\right) \partial _{i}X^{I}& =0\text{ }.  \label{scalar}
\end{align}
As a solution we start with the following metric 
\begin{equation}
ds^{2}=\epsilon _{0}d\tau ^{2}+g_{\mu \nu }dx^{\mu }dx^{\nu }=\epsilon
_{0}d\tau ^{2}+g_{zz}dz^{2}+g_{ab}dx^{a}dx^{b}  \label{generic}
\end{equation}
where $\epsilon _{0}=\pm 1$ and $a,$ $b=1,2,3$. The space-time metric, the
gauge and scalar fields are all functions of $\tau $ only. For the gauge
fields, we first consider solutions with only $\mathcal{F}_{\tau z}^{I}$
non-vanishing$.$ In this case, the equations of motion (\ref{maxwell}) give
the solutions 
\begin{equation}
G_{IJ}\mathcal{F}^{J\tau z}=\frac{q_{I}}{\omega }  \label{ge}
\end{equation}
where $\omega =\sqrt{|g|}$ and $q_{I}$ are constants.

Using the variables introduced in \cite{landau} 
\begin{equation}
\kappa _{\mu \nu }=\dot{g}_{\mu \nu }\text{ },\text{ \ \ }\kappa ^{\mu
}{}_{\nu }=g^{\mu \rho }\kappa _{\rho \nu }\text{ },\text{ \ \ }\kappa ^{\mu
}{}_{\mu }=2\frac{\dot{\omega}}{\omega }\text{ },\text{ }  \label{def}
\end{equation}
where the overdot denotes differentiation with respect to $\tau ,$ the
non-vanishing components of the Ricci tensor for the metric (\ref{generic})
can be written as 
\begin{equation}
R^{\tau }{}_{\tau }=-\epsilon _{0}\left( \frac{\ddot{\omega}}{\omega }-\frac{
\dot{\omega}^{2}}{\omega ^{2}}+\frac{1}{4}\kappa ^{\mu }{}_{\nu }\kappa
^{\nu }{}_{\mu }\right) \text{ },\text{ \ \ \ \ \ }R^{\mu }{}_{\nu }=-\frac{
\epsilon _{0}}{2}\left( \dot{\kappa}^{\mu }{}_{\nu }+\frac{\dot{\omega}}{
\omega }\kappa ^{\mu }{}_{\nu }\right) \text{ }.
\end{equation}
The Einstein equations of motion (\ref{einstein}) then give 
\begin{align}
\frac{\ddot{\omega}}{\omega }-\frac{\dot{\omega}^{2}}{\omega ^{2}}+\frac{1}{
4 }\kappa ^{\mu }{}_{\nu }\kappa ^{\nu }{}_{\mu }+G_{IJ}\dot{X}^{I}\dot{X}
^{J}+ \frac{2\varepsilon }{3\omega ^{2}}g_{zz}G^{IJ}q_{I}q_{J}\text{{\ }}& =0
\text{ },  \label{b} \\
\dot{\kappa}^{z}{}_{z}+\frac{\dot{\omega}}{\omega }\kappa ^{z}{}_{z}+{\frac{
4\varepsilon }{3\omega ^{2}}}g_{zz}G^{IJ}q_{I}q_{J}{\ }& {=0}\text{ }{,}
\label{bb} \\
\dot{\kappa}^{a}{}_{b}+\frac{\dot{\omega}}{\omega }\kappa ^{a}{}_{b}-{\frac{
2\varepsilon }{3\omega ^{2}}}g_{zz}G^{IJ}q_{I}q_{J}\delta ^{a}{}_{b}& =0 
\text{ }.  \label{bbb}
\end{align}
Taking the trace of (\ref{bbb}) and adding (\ref{bb}) we obtain 
\begin{equation}
\ddot{\omega}=\frac{\varepsilon }{3\omega }g_{zz}G^{IJ}q_{I}q_{J}\text{{\ }}.
\label{mb}
\end{equation}
Eq. (\ref{mb}) when substituted back into Eqs. (\ref{b}), (\ref{bb}) and ( 
\ref{bbb}) result in 
\begin{align}
3\frac{\ddot{\omega}}{\omega }-\frac{\dot{\omega}^{2}}{\omega ^{2}}+\frac{1}{
4}\kappa ^{\mu }{}_{\nu }\kappa ^{\nu }{}_{\mu }+G_{IJ}\dot{X}^{I}\dot{X}
^{J}& =0\text{ },  \label{go} \\
\dot{\kappa}^{z}{}_{z}+\frac{\dot{\omega}}{\omega }\kappa ^{z}{}_{z}+4\frac{ 
\ddot{\omega}}{\omega }& =0\text{ }, \\
\dot{\kappa}^{a}{}_{b}+\frac{\dot{\omega}}{\omega }\kappa ^{a}{}_{b}-2\frac{ 
\ddot{\omega}}{\omega }\delta ^{a}{}_{b}& =0\text{ }.
\end{align}
The last two equations can be integrated to give 
\begin{align}
\kappa ^{z}{}_{z}& =\frac{1}{\omega }\left( \theta ^{z}{}_{z}-4\dot{\omega}
\right) \text{ },  \notag \\
\kappa ^{a}{}_{b}& =\frac{1}{\omega }\left( \theta ^{a}{}_{b}+2\dot{\omega}
\delta ^{a}{}_{b}\right) \text{ },  \label{sl2}
\end{align}
where $\theta ^{z}{}_{z}$ and $\theta ^{a}{}_{b}$ are constants satisfying
the condition 
\begin{equation}
\theta ^{z}{}_{z}+\theta ^{a}{}_{a}=0\text{ .}  \label{m2}
\end{equation}
For the metric components we obtain 
\begin{equation}
\dot{g}_{zz}=\frac{1}{\omega }\left( \theta ^{z}{}_{z}{-4}\dot{\omega}
\right) g_{zz}\text{ },\text{ \ \ \ \ \ \ }\dot{g}_{ab}=\frac{1}{\omega }
\left( g_{ac}\theta ^{c}{}_{b}+2\dot{\omega}g_{ab}\right) \text{ .}
\label{metricone}
\end{equation}
Substituting (\ref{sl2}) back into (\ref{go}), we obtain 
\begin{equation}
\frac{1}{4}\theta ^{\mu }{}_{\nu }\theta ^{\nu }{}_{\mu }+3\left( 2\dot{
\omega}^{2}-\theta ^{z}{}_{z}\dot{\omega}+\omega \ddot{\omega}\right)
+\omega ^{2}G_{IJ}\dot{X}^{I}\dot{X}^{J}=0\text{ }.  \label{generalcon}
\end{equation}
The scalar equation of motion (\ref{scalar}), for our solution, using very
special geometry relations, reduces to 
\begin{equation}
\partial _{i}G^{IJ}\left[ \dot{\omega}X_{J}\dot{X}_{I}+\omega \left( X_{J} 
\ddot{X}_{I}-\dot{X}_{I}\dot{X}_{J}\right) -\frac{4}{9}\frac{\varepsilon }{
\omega }g_{zz}q_{I}q_{J}\right] =0\text{ }.  \label{net}
\end{equation}
We conclude that our solutions are defined by Eqs. (\ref{mb}), (\ref{m2}), ( 
\ref{metricone}), (\ref{generalcon}) and (\ref{net}).

We also consider solutions with space-time metric 
\begin{equation}
ds^{2}=\epsilon _{0}d\tau ^{2}+g_{\mu \nu }dx^{\mu }dx^{\nu }=\epsilon
_{0}d\tau ^{2}+g_{xx}dx^{2}+g_{yy}dy^{2}+g_{ab}dx^{a}dx^{b}\text{ }
\end{equation}
where $\epsilon _{0}=\pm 1$ and $a,$ $b=1,2$ and with gauge fields given by 
\begin{equation}
\mathcal{F}_{xy}^{I}=p^{I}\text{ }.
\end{equation}
Here $p^{I}$ are constants and the metric and scalar fields depend only on $
\tau $.

The analysis of the equations of motion (\ref{einstein}) yields 
\begin{align}
\frac{3\ddot{\omega}}{2\omega }-\frac{\dot{\omega}^{2}}{\omega ^{2}}+\frac{1 
}{4}\kappa ^{\mu }{}_{\nu }\kappa ^{\nu }{}_{\mu }+G_{IJ}\dot{X}^{I}\dot{X}
^{J}& =0\text{ },  \label{qw1} \\
\dot{\kappa}^{x}{}_{x}+\frac{\dot{\omega}}{\omega }\kappa ^{x}{}_{x}-2\frac{ 
\ddot{\omega}}{\omega }& =0\text{ },  \label{qw2} \\
\dot{\kappa}^{y}{}_{y}+\frac{\dot{\omega}}{\omega }\kappa ^{y}{}_{y}-2\frac{ 
\ddot{\omega}}{\omega }& =0\text{ },\text{ }  \label{qw22} \\
\dot{\kappa}^{a}{}_{b}+\frac{\dot{\omega}}{\omega }\kappa ^{a}{}_{b}+\frac{ 
\ddot{\omega}}{\omega }\delta ^{a}{}_{b}& =0\text{ },  \label{qw3}
\end{align}
together with the relation 
\begin{equation}
\ddot{\omega}=-\frac{2}{3}\varepsilon \epsilon
_{0}g^{xx}g^{yy}G_{IJ}p^{J}p^{I}\omega \text{ }.  \label{trmag}
\end{equation}
Eqs. (\ref{qw2}), (\ref{qw22}) and (\ref{qw3}) can be integrated to give 
\begin{align}
\kappa ^{x}{}_{x}& =\frac{1}{\omega }\left( \theta ^{x}{}_{x}+2\dot{\omega}
\right) \text{ },  \notag \\
\kappa ^{y}{}_{y}& =\frac{1}{\omega }\left( \theta ^{y}{}_{y}+2\dot{\omega}
\right) \text{ },  \notag \\
\kappa ^{a}{}_{b}& =\frac{1}{\omega }\left( \theta ^{a}{}_{b}-\dot{\omega}
\delta ^{a}{}_{b}\right) \text{ },  \label{k2}
\end{align}
with the condition 
\begin{equation}
\theta ^{x}{}_{x}+\theta ^{y}{}_{y}+\theta ^{a}{}_{a}=0\text{ }.  \label{m3}
\end{equation}
The metric components thus satisfy 
\begin{align}
\dot{g}_{xx}& =\frac{1}{\omega }\left( \theta ^{x}{}_{x}+2\dot{\omega}
\right) g_{xx}\text{ },\text{ \ \ \ \ }  \notag \\
\dot{g}_{yy}& =\frac{1}{\omega }\left( \theta ^{y}{}_{y}+2\dot{\omega}
\right) g_{yy}\text{ , \ \ \ \ \ \ }  \notag \\
\dot{g}_{ab}& =\frac{1}{\omega }\left( g_{ac}\theta ^{c}{}_{b}-g_{ab}\dot{
\omega}\right) \text{ }.  \label{met3}
\end{align}
Upon substituting (\ref{k2}) into (\ref{qw1}), we obtain 
\begin{equation}
\frac{1}{2}\theta ^{\mu }{}_{\nu }\theta ^{\nu }{}_{\mu }+3\left( \dot{
\omega }^{2}-\dot{\omega}\theta ^{a}{}_{a}+\omega \ddot{\omega}\right)
+2\omega ^{2}G_{IJ}\dot{X}^{I}\dot{X}^{J}=0\text{ }.  \label{inter}
\end{equation}
The scalar equation for our solution, with the help of very special geometry
relation, takes the form 
\begin{equation}
\partial _{i}G_{IJ}\left( \frac{\dot{\omega}}{\omega }\dot{X}^{J}X^{I}+X^{I} 
\ddot{X}^{J}-\dot{X}^{I}\dot{X}^{J}+\epsilon _{0}\varepsilon
p^{I}p^{J}g^{xx}g^{yy}\right) =0\text{ .}  \label{mg}
\end{equation}
Eqs (\ref{trmag}), (\ref{m3}), (\ref{met3}), (\ref{inter}) and (\ref{mg})
completely fix this class of solutions.

\section{FAMILIES OF SOLUTIONS}

In what follows, we will construct four families of solutions by solving the
equations characterizing the two classes of solutions analyzed in the
previous section. The first family of solutions is obtained by solving (\ref
{mb}), (\ref{metricone}), (\ref{generalcon}) and (\ref{net}) using the
following change of variables \cite{new} 
\begin{equation}
\frac{d\sigma }{d\tau }=\frac{1}{H(\sigma )}\text{ },\text{ \ \ \ \ \ \ \ \ }
\omega =\sigma H(\sigma )\text{ }.  \label{cv}
\end{equation}
Defining the convenient parameters 
\begin{equation}
\lambda ^{z}{}_{z}=p=\frac{1}{2}\left( \theta ^{z}{}_{z}-4\right) \text{ },
\text{ \ \ }\lambda ^{a}{}_{b}=\frac{1}{2}\left( \theta ^{a}{}_{b}+2\delta
^{a}{}_{b}\right) \text{ },  \label{cov}
\end{equation}
then the analysis of (\ref{metricone}) gives the following metric solution 
\begin{equation}
ds^{2}=H^{2}\left( \epsilon _{0}d\sigma ^{2}+h_{ac}\left( e^{2\lambda \log
\sigma }\right) ^{c}{}_{b}dx^{a}dx^{b}\right) +h_{zz}\sigma ^{2p}H^{-4}dz^{2}
\label{one}
\end{equation}
with $h_{zz}=\pm 1$ and $h_{ac}$ are constants$.$ Note that the condition ( 
\ref{m2}) in terms of the new parameters becomes 
\begin{equation}
p+\lambda ^{a}{}_{a}=1\text{ .}  \label{di}
\end{equation}
Eq. (\ref{generalcon}), using variables (\ref{cv}) and (\ref{cov}), gives 
\begin{equation}
G_{IJ}\frac{dX^{I}}{d\sigma }\frac{dX^{J}}{d\sigma }=\frac{1}{\sigma ^{2}}
\left( 1-\lambda ^{\mu }{}_{\nu }\lambda ^{\nu }{}_{\mu }\right) -\frac{3}{
H^{2}}\left[ \left( \frac{dH}{d\sigma }\right) ^{2}+\frac{\left( 1-2p\right) 
}{\sigma }H\frac{dH}{d\sigma }+H\frac{d^{2}H}{d\sigma ^{2}}\right] 
\label{fii}
\end{equation}
and we obtain from (\ref{mb}) the condition 
\begin{equation}
G^{IJ}q_{I}q_{J}=3\varepsilon h_{zz}\sigma ^{1-2p}H^{3}\left[ \frac{dH}{
d\sigma }+\sigma \frac{d^{2}H}{d\sigma ^{2}}-\frac{\sigma }{H}\left( \frac{dH
}{d\sigma }\right) ^{2}\text{ }\right] \text{ }.  \label{2nd}
\end{equation}
To proceed, we write 
\begin{equation}
X_{I}=\frac{1}{3H^{2}}f_{I}\left( \sigma \right) \text{ },  \label{sp}
\end{equation}
then with the help of very special geometry relations, the following
equations can be derived 
\begin{align}
\frac{dH}{d\sigma }& =\frac{1}{6H}X^{I}\text{ }\frac{df_{I}}{d\sigma }\text{ 
},  \notag \\
H\frac{d^{2}H}{d\sigma ^{2}}-\left( \frac{dH}{d\sigma }\right) ^{2}& =\frac{1
}{6}X^{I}\frac{d^{2}f_{I}}{d\sigma ^{2}}-\frac{1}{12H^{2}}G^{IJ}\frac{df_{I}
}{d\sigma }\frac{df_{J}}{d\sigma }\text{ },  \notag \\
G_{IJ}\frac{dX^{I}}{d\sigma }\frac{dX^{J}}{d\sigma }& =-\frac{3}{H}\frac{
d^{2}H}{d\sigma ^{2}}-\frac{3}{H^{2}}\left( \frac{dH}{d\sigma }\right) ^{2}+
\frac{1}{2H^{2}}X^{I}\frac{d^{2}f_{I}}{d\sigma ^{2}}\text{ }.
\end{align}
Using these relations in (\ref{fii}), we obtain 
\begin{equation}
X^{I}\left( \frac{d^{2}f_{I}}{d\sigma ^{2}}+\frac{\left( 1-2p\right) }{
\sigma }\frac{df_{I}}{d\sigma }+\frac{2}{3\sigma ^{2}}\left( \lambda ^{\mu
}{}_{\nu }\lambda ^{\nu }{}_{\mu }-1\right) f_{I}\right) =0\text{ }.
\end{equation}
This admits the solutions 
\begin{equation}
f_{I}=A_{I}\sigma ^{\left( p-s\right) }+B_{I}\sigma ^{\left( p+s\right) }
\label{sol}
\end{equation}
with 
\begin{equation}
\lambda ^{\mu }{}_{\nu }\lambda ^{\nu }{}_{\mu }=1+\frac{3}{2}\left(
p^{2}-s^{2}\right)   \label{dis}
\end{equation}
and constant $A_{I}$ and $B_{I}.$

We also obtain from (\ref{2nd}) the condition 
\begin{equation}
G^{IJ}\left( f_{J}\frac{d^{2}f_{I}}{d\sigma ^{2}}+\frac{1}{\sigma }f_{J} 
\frac{df_{I}}{d\sigma }-\frac{df_{I}}{d\sigma }\frac{df_{J}}{d\sigma }
-4\varepsilon h_{zz}q_{I}q_{J}\sigma ^{2p-2}\right) =0\text{ }.  \label{fee}
\end{equation}
Substituting our solution into (\ref{fee}), gives the algebraic condition 
\begin{equation}
G^{IJ}\left( s^{2}A_{J}B_{I}-\varepsilon h_{zz}q_{I}q_{J}\right) =0\text{ }.
\label{conone}
\end{equation}
Eq. (\ref{net}) for our solutions, after some calculations, reduces to 
\begin{equation}
\partial _{i}G^{IJ}\left( s^{2}A_{J}B_{I}-\varepsilon
h_{zz}q_{I}q_{J}\right) =0\text{ }.  \label{contwo}
\end{equation}
Finally Eqs. (\ref{cn}) and (\ref{vs}) imply 
\begin{equation}
H^{4}=\frac{1}{6}G^{IJ}f_{I}f_{J}\text{ }.  \label{wa}
\end{equation}
Note that for $s=p$ and diagonal $\lambda ^{\mu }{}_{\nu }$, we reproduce
the solutions obtained in \cite{s2}.

We can also get a second family of solutions by making the following change
of variables 
\begin{equation}
\frac{d\rho }{d\tau }=\frac{1}{\omega (\rho )}\text{ }.\text{ }  \label{cvv}
\end{equation}
Starting with Eq. (\ref{metricone}), we obtain the metric solution 
\begin{equation}
ds^{2}=\omega ^{2}\left( \epsilon _{0}d\rho ^{2}+h_{ac}\left( e^{\theta \rho
}\right) ^{c}{}_{b}dx^{a}dx^{b}\right) +\frac{1}{\omega ^{4}}h_{zz}e^{\theta
^{z}{}_{z}\rho }dz^{2}\text{ }.  \label{two}
\end{equation}
with $h_{zz}=\pm 1$, $h_{ac}$ are constants and 
\begin{equation}
\theta ^{z}{}_{z}+\theta ^{a}{}_{a}=0.
\end{equation}
To proceed, we write 
\begin{align}
X_{I}& =\frac{1}{3\omega ^{2}}g_{I}\left( \rho \right) \text{ }, \\
\omega ^{4}& =\frac{1}{6}G^{IJ}g_{I}g_{J}\text{ }.
\end{align}
Using the relations of very special geometry, then (\ref{generalcon}), (\ref
{mb}) and (\ref{net}), respectively, give 
\begin{align}
X^{I}\left( \frac{d^{2}g_{I}}{d\rho ^{2}}-\theta ^{z}{}_{z}\frac{dg_{I}}{
d\rho }+\frac{1}{6}\theta ^{\mu }{}_{\nu }\theta ^{\nu }{}_{\mu
}g_{I}\right) & =0\text{ },  \label{echo} \\
G^{IJ}\left( g_{J}\frac{d^{2}g_{I}}{d\rho ^{2}}-\frac{dg_{I}}{d\rho }\frac{
dg_{J}}{d\rho }-4\varepsilon h_{zz}e^{\theta ^{z}{}_{z}\rho
}q_{I}q_{J}\right) & =0\text{ },  \label{bet} \\
\partial _{i}G^{IJ}\left( g_{J}\frac{d^{2}g_{I}}{d\rho ^{2}}-\frac{dg_{I}}{
d\rho }\frac{dg_{J}}{d\rho }-4\varepsilon h_{zz}e^{\theta ^{z}{}_{z}\rho
}q_{I}q_{J}\right) & =0\text{ }.  \label{pink}
\end{align}
Eq. (\ref{echo}) admits the solution 
\begin{equation}
g_{I}=A_{I}e^{\frac{1}{2}\left( \theta ^{z}{}_{z}-s\right) \rho }+B_{I}e^{
\frac{1}{2}\left( \theta ^{z}{}_{z}+s\right) \rho }
\end{equation}
$\allowbreak $ with 
\begin{equation}
\theta ^{a}{}_{b}\theta ^{b}{}_{a}=\frac{1}{2}\left( \theta
^{z}{}_{z}\right) ^{2}-\frac{3}{2}s^{2}\text{ }.
\end{equation}
This solution when substituted in (\ref{bet}) and (\ref{pink}) gives the
algebraic conditions 
\begin{equation}
G^{IJ}\left( s^{2}A_{I}B_{J}-4\varepsilon h_{zz}q_{I}q_{J}\right) =\partial
_{i}G^{IJ}\left( s^{2}A_{I}B_{J}-4\varepsilon h_{zz}q_{I}q_{J}\right) =0
\text{ }.
\end{equation}

A third family of solutions is obtained by solving (\ref{trmag}), (\ref{met3}
), (\ref{inter}) and (\ref{mg}) using the change of variables given by (\ref
{cv}). This gives for the metric solution
\begin{equation}
ds^{2}=H^{2}\left( \epsilon _{0}d\tau ^{2}+h_{xx}\sigma
^{2p}dx^{2}+h_{yy}\sigma ^{2r}dy^{2}\right) +h_{ac}\left( e^{2\lambda \log
\sigma }\right) ^{c}{}_{b}\frac{1}{H}dx^{a}dx^{b}\text{ }.  \label{three}
\end{equation}
where we have defined 
\begin{equation}
\lambda ^{x}{}_{x}=p=\frac{1}{2}\left( \theta _{\text{ \ }x}^{x}+2\right) ,\
\ \ \ \lambda ^{y}{}_{y}=r=\frac{1}{2}\left( \theta _{\text{ \ }
y}^{y}+2\right) ,\text{ \ \ \ }\lambda ^{a}{}_{b}=\frac{1}{2}\left( \theta
^{a}{}_{b}-\delta ^{a}{}_{b}\right) \text{ }.
\end{equation}
and $h_{xx}$, $h_{yy}$ take the values $\pm 1$ and $h_{ac}$ are constants.
The condition (\ref{m3}) becomes 
\begin{equation}
p+r+\lambda ^{a}{}_{a}=1\text{ }.
\end{equation}
To proceed, we set 
\begin{equation}
X^{I}=\frac{1}{H}h^{I}(\sigma ).
\end{equation}
Again with the help of very special geometry, we obtain the following useful
relations 
\begin{align}
\frac{dH}{d\sigma }& =X_{I}\frac{dh^{I}}{d\sigma }\text{ }, \\
-\frac{d^{2}H}{d\sigma ^{2}}+\frac{1}{H}\left( \frac{dH}{d\sigma }\right)
^{2}+X_{I}\frac{d^{2}h^{I}}{d\sigma ^{2}}& =\frac{2}{3H}G_{IJ}\frac{dh^{I}}{
d\sigma }\frac{dh^{J}}{d\sigma }\text{ }, \\
G_{IJ}\frac{dX^{I}}{d\sigma }\frac{dX^{J}}{d\sigma }& =\frac{3}{2H}\left( -
\frac{d^{2}H}{d\sigma ^{2}}+X_{I}\frac{d^{2}h^{I}}{d\sigma ^{2}}\right) 
\text{ }.
\end{align}
These relations, when used in Eqs. (\ref{inter}), (\ref{trmag}) and (\ref{mg}
), give 
\begin{equation}
X_{I}\left[ \frac{2}{3}\left( \lambda ^{\mu }{}_{\nu }\lambda ^{\nu }{}_{\mu
}-1\right) h^{I}+\left( 1-2l\right) \sigma \frac{dh^{I}}{d\sigma }+\sigma
^{2}\frac{d^{2}h^{I}}{d\sigma ^{2}}\right] =0\text{ ,}  \label{vera}
\end{equation}
and 
\begin{eqnarray}
G_{IJ}\left[ h^{J}\left( \frac{1}{\sigma }\frac{dh^{I}}{d\sigma }+\frac{
d^{2}h^{I}}{d\sigma ^{2}}\right) -\frac{dh^{I}}{d\sigma }\frac{dh^{J}}{
d\sigma }+\sigma ^{2l-2}\varepsilon \epsilon _{0}h_{xx}h_{yy}p^{J}p^{I}
\right]  &=&0\text{ },  \label{al} \\
\partial _{i}G_{IJ}\left[ h^{J}\left( \frac{1}{\sigma }\frac{dh^{I}}{d\sigma 
}+\frac{d^{2}h^{I}}{d\sigma ^{2}}\right) -\frac{dh^{I}}{d\sigma }\frac{dh^{J}
}{d\sigma }+\sigma ^{2l-2}\varepsilon \epsilon _{0}h_{xx}h_{yy}p^{J}p^{I}
\right]  &=&0\text{ }.  \label{al1}
\end{eqnarray}
where we have defined $l=\lambda ^{a}{}_{a}.$ Eq. (\ref{vera}) admits the
solution 
\begin{equation}
h^{I}=C^{I}\sigma ^{l-s}+D^{I}\sigma ^{l+s}\text{ },
\end{equation}
with constant $C^{I}$ and $D^{I}$ and 
\begin{equation*}
\lambda ^{\mu }{}_{\nu }\lambda ^{\nu }{}_{\mu }=p^{2}+r^{2}+\lambda
^{a}{}_{b}\lambda ^{b}{}_{a}=1+\frac{3}{2}(l^{2}-s^{2})\text{ }.
\end{equation*}
Substituting this solution into (\ref{al}) and (\ref{al1}), we obtain the
algebraic conditions 
\begin{equation}
G_{IJ}\left( 4s^{2}C^{I}D^{J}+\varepsilon \epsilon _{0}h_{xx}h_{yy}p^{J}p^{I}
\text{ }\right) =\partial _{i}G_{IJ}\left( 4s^{2}C^{I}D^{J}+\varepsilon
\epsilon _{0}h_{xx}h_{yy}p^{J}p^{I}\right) =0\text{ }.
\end{equation}
Employing the condition (\ref{cn}) fixes $H$ and we obtain 
\begin{equation}
H^{3}={\frac{1}{6}}C_{IJK}h^{I}h^{J}h^{K}\text{ }.
\end{equation}

Finally, a fourth family of solutions can be obtained by solving (\ref{trmag}
), (\ref{met3}), (\ref{inter}) and (\ref{mg}) using the change of variables
given by (\ref{cvv}). Solving for the metric we obtain 
\begin{equation}
ds^{2}=\omega ^{2}\left( \epsilon _{0}d\rho ^{2}+h_{xx}e^{\theta
^{x}{}_{x}\rho }dx^{2}+h_{yy}e^{\theta ^{y}{}_{y}\rho }dy^{2}\right) +h_{ac}
\frac{1}{\omega }\left( e^{\theta \rho }\right) ^{c}{}_{b}dx^{a}dx^{b}\text{ 
}.  \label{four}
\end{equation}
with $h_{xx}$ and $h_{yy}$ taking the values $\pm 1$, $h_{ac}$ are constants
and 
\begin{equation}
\theta ^{x}{}_{x}+\theta ^{y}{}_{y}+\theta ^{a}{}_{a}=0\text{ }.
\end{equation}
To proceed we write 
\begin{eqnarray}
X^{I} &=&\frac{1}{\omega }h^{I}\left( \rho \right) \text{ }, \\
\omega ^{3} &=&{\frac{1}{6}}C_{IJK}h^{I}h^{J}h^{K}\text{ }.
\end{eqnarray}
We obtain, after some calculation using very special geometry relations, the
following equations 
\begin{eqnarray}
X_{I}\text{ }\left[ \frac{1}{2}\theta ^{\mu }{}_{\nu }\theta ^{\nu }{}_{\mu
}h^{I}+3\left( \frac{d^{2}h^{I}}{d\sigma ^{2}}-l\frac{dh^{I}}{d\sigma }
\right) \right]  &=&0\text{ },  \label{kar1} \\
G_{IJ}\left( h^{J}\frac{d^{2}h^{I}}{d\sigma ^{2}}-\frac{dh^{I}}{d\sigma }
\frac{dh^{J}}{d\sigma }+\varepsilon \epsilon _{0}h_{xx}h_{yy}e^{-\rho \left(
\theta ^{x}{}_{x}+\theta ^{y}{}_{y}\right) }p^{J}p^{I}\text{ }\right)  &=&0
\text{ },  \label{ka} \\
\partial _{i}G_{IJ}\left( h^{J}\frac{d^{2}h^{I}}{d\sigma ^{2}}-\frac{dh^{I}}{
d\sigma }\frac{dh^{J}}{d\sigma }+\varepsilon \epsilon
_{0}h_{xx}h_{yy}e^{-\rho \left( \theta ^{x}{}_{x}+\theta ^{y}{}_{y}\right)
}p^{J}p^{I}\text{ }\right)  &=&0\text{ }.  \label{ka1}
\end{eqnarray}
with $l=\theta ^{a}{}_{a}.$ Eq. (\ref{kar1}) can be solved by 
\begin{equation}
h^{I}=C^{I}e^{\frac{1}{2}\left( l-s\right) \rho }+D^{I}e^{\frac{1}{2}\left(
l+s\right) \rho }\allowbreak 
\end{equation}
with 
\begin{equation}
\theta ^{\mu }{}_{\nu }\theta ^{\nu }{}_{\mu }=\left( \theta
^{x}{}_{x}\right) ^{2}+\left( \theta ^{y}{}_{y}\right) ^{2}+\theta
^{a}{}_{b}\theta ^{b}{}_{a}=\frac{3}{2}\left( l^{2}-s^{2}\right) \text{ }.
\end{equation}
Upon substituting the solution in (\ref{ka}) and (\ref{ka1}), we obtain the
algebraic conditions 
\begin{equation}
G_{IJ}\left( s^{2}C^{I}D^{I}+\varepsilon \epsilon
_{0}h_{xx}h_{yy}p^{J}p^{I}\right) =\partial _{i}G_{IJ}\left(
s^{2}C^{I}D^{I}+\varepsilon \epsilon _{0}h_{xx}h_{yy}p^{J}p^{I}\right) =0
\text{ }.
\end{equation}

\section{EXAMPLES}

We construct solutions with three charges for the STU supergravity model
with the prepotential $\mathcal{V}=X^{1}X^{2}X^{3}=1$. This theory can be
obtained via the reduction of eleven-dimensional supergravity on $T^{6}$.
For this model, the gauge coupling matrix with element $G_{IJ},$ is given by 
\begin{equation}
G=\frac{1}{2}\left( 
\begin{array}{ccc}
\left( X^{1}\right) ^{-2} & 0 & 0 \\ 
0 & \left( X^{2}\right) ^{-2} & 0 \\ 
0 & 0 & \left( X^{3}\right) ^{-2}
\end{array}
\right) \text{ }.
\end{equation}
We start with the first family of solutions where the metric is given by ( 
\ref{one})$.$ Using (\ref{sp}) we obtain the following solutions 
\begin{align}
X^{2}X^{3}& =\frac{1}{H^{2}}f_{1}\text{ },\text{ \ \ \ \ \ }  \notag \\
X^{1}X^{3}& =\frac{1}{H^{2}}f_{2}\text{ },\text{ \ \ \ \ \ \ }  \notag \\
X^{1}X^{2}& =\frac{1}{H^{2}}f_{3}\text{ },
\end{align}
with 
\begin{equation}
f_{I}=\sigma ^{\left( p-s\right) }\left( 1+B_{I}\sigma ^{2s}\right) \text{ }.
\end{equation}
The scalars and gauge fields are given by 
\begin{align}
X^{1}& =\frac{\left( f_{1}f_{2}f_{3}\right) ^{\frac{1}{3}}}{f_{1}},\text{ \
\ \ \ \ \ \ \ \ \ \ \ \ }X^{2}=\frac{\left( f_{1}f_{2}f_{3}\right) ^{\frac{1 
}{3}}}{f_{2}},\text{ \ \ \ \ \ \ \ \ \ \ \ \ \ }X^{3}=\frac{\left(
f_{1}f_{2}f_{3}\right) ^{\frac{1}{3}}}{f_{3}}\text{ },  \notag \\
\mathcal{F}^{1\sigma z}& =2q_{1}\frac{\left( f_{1}f_{2}f_{3}\right) ^{\frac{
1 }{3}}}{\sigma f_{1}^{2}}.\text{ \ \ \ \ \ \ }\mathcal{F}^{2\sigma
z}=2q_{2} \frac{\left( f_{1}f_{2}f_{3}\right) ^{\frac{1}{3}}}{\sigma
f_{2}^{2}},\text{ \ \ \ \ \ }\mathcal{F}^{3\sigma z}=2q_{3}\frac{\left(
f_{1}f_{2}f_{3}\right) ^{\frac{1}{3}}}{\sigma f_{3}^{2}}\text{ }.
\end{align}
Our solutions have three independent charges $q_{I}$ with 
\begin{equation}
B_{I}=\frac{1}{s^{2}}\varepsilon h_{zz}q_{I}^{2}\text{ .}
\end{equation}
The metric, using (\ref{wa})$,$ is given by 
\begin{equation}
ds^{2}=\left( f_{1}f_{2}f_{3}\right) ^{\frac{1}{3}}\left( \epsilon
_{0}d\sigma ^{2}+h_{ac}\left( e^{2\lambda \log \sigma }\right)
^{c}{}_{b}dx^{a}dx^{b}\right) +h_{zz}\sigma ^{2p}\frac{1}{\left(
f_{1}f_{2}f_{3}\right) ^{\frac{2}{3}}}dz^{2}
\end{equation}
with the conditions 
\begin{align}
\lambda ^{a}{}_{a}& =1-p\text{ },  \notag \\
\lambda ^{a}{}_{b}\lambda ^{b}{}_{a}& =1-\frac{3}{2}s^{2}+\frac{1}{2}p^{2} 
\text{ }.
\end{align}

For the second family of solutions described by the metric (\ref{two}), we
obtain 
\begin{equation}
ds^{2}=\left( g_{1}g_{2}g_{3}\right) ^{\frac{1}{3}}\left( \epsilon _{0}d\rho
^{2}+h_{ac}\left( e^{\theta \rho }\right) ^{c}{}_{b}dx^{a}dx^{b}\right) + 
\frac{1}{\left( g_{1}g_{2}g_{3}\right) ^{\frac{2}{3}}}h_{zz}e^{\theta
^{z}{}_{z}\rho }dz^{2}\text{ }.
\end{equation}
and 
\begin{align}
X^{1}& =\frac{\left( g_{1}g_{2}g_{3}\right) ^{\frac{1}{3}}}{g_{1}},\text{ \
\ \ \ \ \ \ \ \ }X^{2}=\frac{\left( g_{1}g_{2}g_{3}\right) ^{\frac{1}{3}}}{
g_{2}},\text{ \ \ \ \ \ \ \ \ \ \ \ \ \ \ \ \ }X^{3}=\frac{\left(
g_{1}g_{2}g_{3}\right) ^{\frac{1}{3}}}{g_{3}}\text{ },\text{ }  \notag \\
\mathcal{F}^{1\rho z}& =2q_{1}\frac{\left( g_{1}g_{2}g_{3}\right) ^{\frac{1}{
3}}}{g_{1}^{2}},\text{\ \ \ \ }\mathcal{F}^{2\rho z}=2q_{2}\frac{\left(
g_{1}g_{2}g_{3}\right) ^{\frac{1}{3}}}{g_{2}^{2}},\text{ \ \ \ \ \ \ \ }
\mathcal{F}^{3\rho z}=2q_{3}\frac{\left( g_{1}g_{2}g_{3}\right) ^{\frac{1}{3}
}}{g_{3}^{2}}
\end{align}
with 
\begin{equation}
g_{I}=e^{\frac{1}{2}\left( \theta ^{z}{}_{z}-s\right) \rho }\left( 1+\frac{
4\varepsilon h_{zz}\left( q_{I}\right) ^{2}}{s^{2}}e^{s\rho }\right) \text{ }
\end{equation}
$\allowbreak $and 
\begin{equation}
\theta ^{z}{}_{z}+\theta ^{a}{}_{a}=0,\text{ \ \ }\theta ^{a}{}_{b}\theta
^{b}{}_{a}=\frac{1}{2}\left( \theta ^{z}{}_{z}\right) ^{2}-\frac{3}{2}s^{2} 
\text{ }.
\end{equation}
Similarly one can construct solutions of the third family with $\mathcal{F}
_{xy}^{I}=p^{I}$ and obtain 
\begin{align}
ds^{2}& =\left( h_{1}h_{2}h_{3}\right) ^{2}\left( \epsilon _{0}d\tau
^{2}+h_{xx}\sigma ^{2p}dx^{2}+h_{yy}\sigma ^{2r}dy^{2}\right) +h_{ac}\left(
e^{2\lambda \log \sigma }\right) ^{c}{}_{b}\frac{1}{\left(
h_{1}h_{2}h_{3}\right) }dx^{a}dx^{b}\text{ },  \notag \\
X^{1}& =\frac{h^{1}}{\left( h^{1}h^{2}h^{3}\right) ^{\frac{1}{3}}}\text{ }, 
\text{ \ }X^{2}=\frac{h^{2}}{\left( h^{1}h^{2}h^{3}\right) ^{\frac{1}{3}}} 
\text{ },\text{ \ \ }X^{3}=\frac{h^{3}}{\left( h^{1}h^{2}h^{3}\right) ^{ 
\frac{1}{3}}}\text{ },  \notag \\
h^{I}& =\sigma ^{l-s}\left( 1-\frac{1}{4s^{2}}\varepsilon \epsilon
_{0}h_{xx}h_{yy}\left( p^{I}\right) ^{2}\sigma ^{2s}\right) \text{ },
\end{align}
with 
\begin{equation}
p+r+\lambda ^{a}{}_{a}=1\text{ },\text{ \ \ \ \ }l=\lambda ^{a}{}_{a}\text{ }
,\text{ \ \ \ \ }p^{2}+r^{2}+\lambda ^{a}{}_{b}\lambda ^{b}{}_{a}=1+\frac{3}{
2}\left( l^{2}-s^{2}\right) \text{ }.
\end{equation}
For the fourth family of solutions we obtain 
\begin{align}
ds^{2}& =\left( h_{1}h_{2}h_{3}\right) ^{2}\left( \epsilon _{0}d\rho
^{2}+h_{xx}e^{\theta ^{x}{}_{x}\rho }dx^{2}+h_{yy}e^{\theta ^{y}{}_{y}\rho
}dy^{2}\right) +h_{ac}\frac{1}{\left( h_{1}h_{2}h_{3}\right) ^{2}}\left(
e^{\theta \rho }\right) ^{c}{}_{b}dx^{a}dx^{b}\text{ },  \notag \\
X^{1}& =\frac{h^{1}}{\left( h^{1}h^{2}h^{3}\right) ^{\frac{1}{3}}}\text{ }, 
\text{ \ }X^{2}=\frac{h^{2}}{\left( h^{1}h^{2}h^{3}\right) ^{\frac{1}{3}}} 
\text{ },\text{ \ \ }X^{3}=\frac{h^{3}}{\left( h^{1}h^{2}h^{3}\right) ^{ 
\frac{1}{3}}}\text{ },  \notag \\
h^{I}& =e^{\frac{\rho }{2}\left( l-s\right) }\left( 1-\frac{1}{s^{2}}
\varepsilon \epsilon _{0}h_{xx}h_{yy}\left( p^{I}\right) ^{2}e^{s\rho
}\allowbreak \right)
\end{align}
with 
\begin{equation}
\theta ^{x}{}_{x}+\theta ^{y}{}_{y}+\theta ^{a}{}_{a}=0,\text{ \ \ \ \ }
l=\theta ^{a}{}_{a}\text{ \ },\text{ \ \ \ \ }\left( \theta
^{x}{}_{x}\right) ^{2}+\left( \theta ^{y}{}_{y}\right) ^{2}+\theta
^{a}{}_{b}\theta ^{b}{}_{a}=\frac{3}{2}\left( l^{2}-s^{2}\right) .
\end{equation}

\section{SUMMARY}

Four families of solutions depending only on one coordinate are obtained for 
$N=2$ five-dimensional supergravity theories. These solutions have
non-trivial scalar and gauge fields and are valid for all space-time
signatures. As examples we constructed explicit solutions with three charges
for the so-called STU model. The constructed solutions can be lifted up to
solutions of eleven-dimensional supergravity as well as type IIB
supergravity (see for example section 4.1 of \cite{reduction}).

The various metric solutions are expressed in terms of constrained matrices.
Their explicit form will depend on the various Jordan normal forms of the
constrained matrices appearing in the metric. One can consider the following
choices for $\lambda ^{\mu }{}_{\nu }$ and $\theta ^{\mu }{}_{\nu }$ 
\begin{align}
& \left( 
\begin{array}{cccc}
p & 0 & 0 & 0 \\ 
0 & r & 0 & 0 \\ 
0 & 0 & a & 0 \\ 
0 & 0 & 0 & b
\end{array}
\right) \text{ \ },\text{ \ \ \ \ \ \ }\left( 
\begin{array}{cccc}
p & 0 & 0 & 0 \\ 
0 & r & 0 & 0 \\ 
0 & 0 & a+ib & 0 \\ 
0 & 0 & 0 & a-ib
\end{array}
\right) \text{ \ },  \notag \\
& \left( 
\begin{array}{cccc}
p & 0 & 0 & 0 \\ 
0 & r & 0 & 0 \\ 
0 & 0 & a & 1 \\ 
0 & 0 & 0 & a
\end{array}
\right) \text{ \ },\text{ \ \ \ \ \ \ \ }\left( 
\begin{array}{cccc}
\beta & 0 & 0 & 0 \\ 
0 & \alpha & 1 & 0 \\ 
0 & 0 & \alpha & 1 \\ 
0 & 0 & 0 & \alpha
\end{array}
\right) \text{ \ },
\end{align}
All these choices can be used to construct solutions for the families
represented by (\ref{one}) and (\ref{two}). For the families represented by (
\ref{three}) and (\ref{four}) only the first three choices are valid. In
working out the explicit solutions, one should take into account that the
metric components are symmetric and that $\omega =\sqrt{|g|}.$ This implies
that the matrix with elements $h_{\mu \nu }$ appearing in our solutions is
symmetric and satisfies the condition $h_{\mu \rho }M^{\rho }{}_{\nu
}=h_{\nu \rho }M^{\rho }{}_{\mu }$ where $M$ is either $\lambda $ or $\theta
.$

The results of this paper should be extended to other supergravity theories
and in particular $N=2$ four-dimensional supergravity. An interesting
direction for further investigation is to generalize our results to
supergravity theories with a cosmological constant or a scalar potential.
Solutions of such theories should have applications to (A)dS/CFT
correspondence. We hope to report on this in our future work.

\end{document}